# Multiferroic nature of charge-ordered rare earth manganites

Claudy Rayan Serrao[1,2], A Sundaresan[1] and C N R Rao[1,2*]

[1]Chemistry and Physics of Materials Unit, Jawaharlal Nehru Centre for Advanced Scientific Research, Jakkur P.O., Bangalore –560064, India.

[2]Material Research Centre, Indian Institute of Science, Bangalore -560012, India.

Charge-ordered rare earth manganites $Nd_{0.5}Ca_{0.5}MnO_3$, $La_{0.25}Nd_{0.25}Ca_{0.5}MnO_3$, $Pr_{0.7}Ca_{0.3}MnO_3$ and $Pr_{0.6}Ca_{0.4}MnO_3$ are found to exhibit dielectric constant anomalies around the charge-ordering or the magnetic transition temperatures. Magnetic fields have a marked effect on the dielectric properties, indicating the presence of coupling between the magnetic and electrical order parameters. The observation of magnetoferroelectricity in these manganites is in accord with the recent theoretical predictions of Khomskii and coworkers.

PACS numbers:

    75.47.Lx  75.80.+q  77.22.-d

---

[*] Corresponding author: cnrrao@jncasr.ac.in

There is intense interest in the study of multiferroics in recent years, and a few novel materials with multiferroic properties have indeed been discovered [1-5]. Multiferroics which exhibit coupling between the magnetic and electrical order parameters are of vital interest both academically and technologically. Coexistence of ferroelectricity and magnetism in transition metal oxide-based materials is generally not favoured since magnetism requires d-electrons and ferroelectricity does not. Magnetoelectric effects in many of the materials therefore occur by alternative mechanisms. Thus, magnetoelectric properties of rare earth manganates such as $YMnO_3$, $TbMnO_3$ and $YMn_2O_5$ arise due to the tilting of polyhedra or frustrated magnetism [6-8]. The 6s lone pair plays a crucial role in the magnetoelectric properties of bismuth-based materials such as $BiFeO_3$ and $BiMnO_3$ [9]. Recently, Khomskii and coworkers [10] have pointed out that coupling between magnetic and charge-ordering in charge-ordered and orbital-ordered perovskites can give rise to ferroelectric magnetism. Charge-ordering in the rare earth manganites, $Ln_{1-x}Ca_xMnO3$, (Ln = rare earth) can be site-centered (SCO) or body-centred (BCO). The SCO behaviuor occurs around $x = 0.5$ with a CE-type antiferromagnetic state while BCO can occur around $x = 0.4$ with a possible perpendicular spin structure. There is a report in the literature for the occurrence of a dielectric anomaly in $Pr_{0.6}Ca_{0.4}MnO_3$ around the charge-ordering transition temperature [11]. Although magnetic fields are noted to affect the dielectric properties of the manganites [12], there has been no definitive study of the effect of magnetic fields on the dielectric properties to establish whether there is coupling between the electrical and magnetic order parameters. We have investigated the dielectric properties of $Pr_{0.7}Ca_{0.3}MnO_3$, $Pr_{0.6}Ca_{0.4}MnO_3$ and $Nd_{0.5}Ca_{0.5}MnO_3$ which have comparable radii of the A-site cations, and exhibit charge-ordering in the 220-240 K region and an antiferromagnetic transition in the 130-170 K region. Electrical resistivity of these manganites is not affected significantly on application of magnetic fields upto 4 T.

$Nd_{0.5}Ca_{0.5}MnO_3$ exhibits a charge-ordered transition at 240 K ($T_{CO}$) and an anti-ferromagnetic transition at 140 K ($T_N$) [13]. The dielectric constant of this material increases substantially in the region of $T_{CO}$ and $T_N$ showing a broad maximum, reaching a value of 500 or above. Application of a magnetic field of 3.3 T has a significant effect on the dielectric constant, showing an increase below $T_N$ as shown from figure 1. $Pr_{0.7}Ca_{0.3}MnO_3$ with a $T_{CO}$ of 225 K and $T_N$ of 130 K [14], shows a broad maximum in the dielectric constant around $T_{CO}$ with the value reaching 2000 or greater (figure 2). Application of a magnetic field of 2 T causes a marked increase in the dielectric constant below $T_{CO}$ as can be seen from figure 2. The results in figures 1 and 2 demonstrate that there is coupling between the electrical and magnetic order parameters in both $Nd_{0.5}Ca_{0.5}MnO_3$ and $Pr_{0.7}Ca_{0.3}MnO_3$.

The case of $Pr_{0.6}Ca_{0.4}MnO_3$ is of special interest because of the considerations mentioned earlier. This manganite exhibits a $T_{CO}$ of 240 K and $T_N$ of 170 K [14]. There is an additional magnetic transition around 50 K. The dielectric constant of this material shows a broad maximum around $T_N$ and the value is also high (figure 3). Application of a magnetic field of 2 T, markedly affects the dielectric properties as shown in figure 3. The sign of the magnetocapacitance effect changes with temperature, being negative near $T_N$ and positive near $T_{CO}$. We observe a crossing of the dielectric constant curves around a temperature between $T_{CO}$ and $T_N$. The magnetoelectric behaviour of this manganite is somewhat different from that of the $Nd_{0.5}Ca_{0.5}MnO_3$ or $Pr_{0.7}Ca_{0.3}MnO_3$. The results, however, demonstrate the presence of magnetoelectricity in all the charge-ordered manganites studied by us as predicted by Khomskii and coworkers [10].

The present study of the charge-ordered manganites shows that many of the charge-ordered rare earth manganites are multiferroics showing magnetocapacitance. They also show exhibit certain other features. Thus, the broad maximum in the dielectric constant around $T_{CO}$ or $T_N$ becomes more prominent in the presence of the magnetic field in certain cases. The magnitude and sign of the magnetocapacitance effect depends on the frequency of measurement. It is generally better to employ a relatively high frequency, well above 10k Hz to obtain reliable data.

We have also studied the dielectric properties of a few other charge-ordered rare earth manganites, $La_{0.25}Nd_{0.25}Ca_{0.5}MnO_3$ being one of them. This material exhibits a re-entrant ferromagnetic transition at a temperature lower than the charge-ordered transition. The charge-ordered transition is around 240 K and the ferromagnetic transition is at 150 K [15]. Interestingly, we find maxima in the dielectric constant at both $T_{CO}$ and $T_N$, the one at $T_{CO}$ being is more prominent (figure 4).

It is important to note some of the important characteristics of the charge-ordered manganites studied by us in order to fully understand their ferroic properties. The most important feature is that all these manganites exhibit electronic phase separation at low temperatures (T < $T_{CO}$) [10]. They exhibit a decrease in resistivity on application of large magnetic fields (>4 T) [16,17]. Application of electric fields also causes a significant decrease in the resistivity of the manganites [17,18]. On the application of electric fields, the manganites show magnetic response [19]. Such electric field-induced magnetization may also be taken as evidence for coupling between the electric and magnetic order parameters in the manganites. It is likely that grain boundaries between the different electronic phases have a role in determining the dielectric behaviour. The importance of grain boundaries in giving rise to high dielectric constants has indeed been recognized [20,21]. In spite of the complexity of their electronic structure, the present study shows that the charge-ordered rare earth manganites possess multiferroic and magnetoelectric properties. Clearly, charge-ordering provides a novel route to multiferroic properties specially in the case of the manganites.

**Acknowledgement**:

The authors thank Manu Hegde and Ramakrishna Matte for assistance with some of the experiments.

**Figure captions:**

Figure 1 Temperature variation of the dielectric constant (1M Hz) of $Nd_{0.5}Ca_{0.5}MnO_3$ in the absence and presence of a magnetic field (H = 3.3 T). The inset shows data at 30k Hz.

Figure 2 Temperature variation of the dielectric constant (1M Hz) of $Pr_{0.7}Ca_{0.3}MnO_3$ in the absence and presence of a magnetic field (H = 2 T). The inset shows data at 30k Hz.

Figure 3 Temperature variation of the dielectric constant (1M Hz) of $Pr_{0.6}Ca_{0.4}MnO_3$ in the absence and presence of a magnetic field (H = 2 T). The inset shows data at 30k Hz

Figure 4 Temperature variation of the dielectric constant of $La_{0.25}Nd_{0.25}Ca_{0.5}MnO_3$. The hatched curve contains data taken in the 100k Hz-1M Hz range.

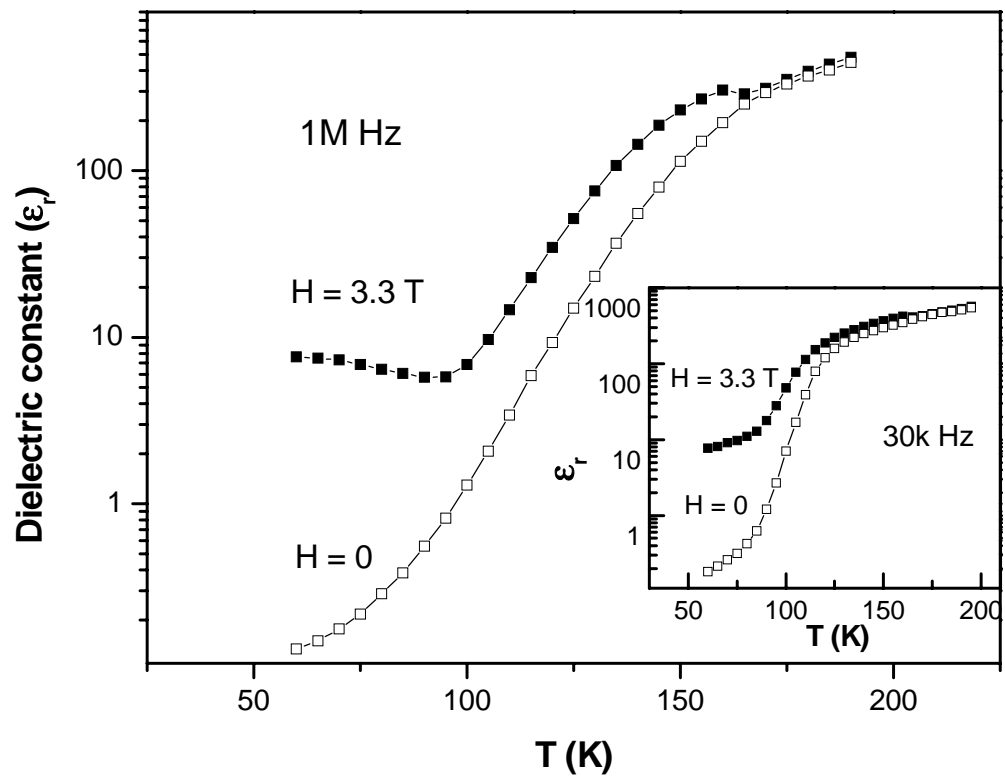

Figure 1

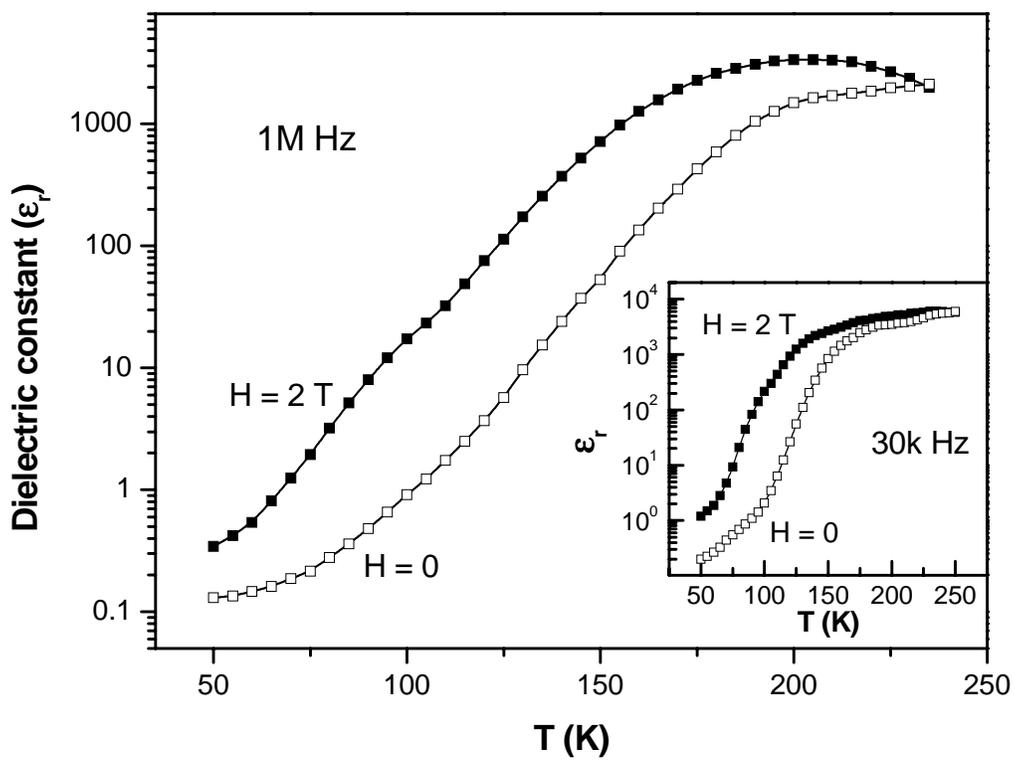

Figure 2

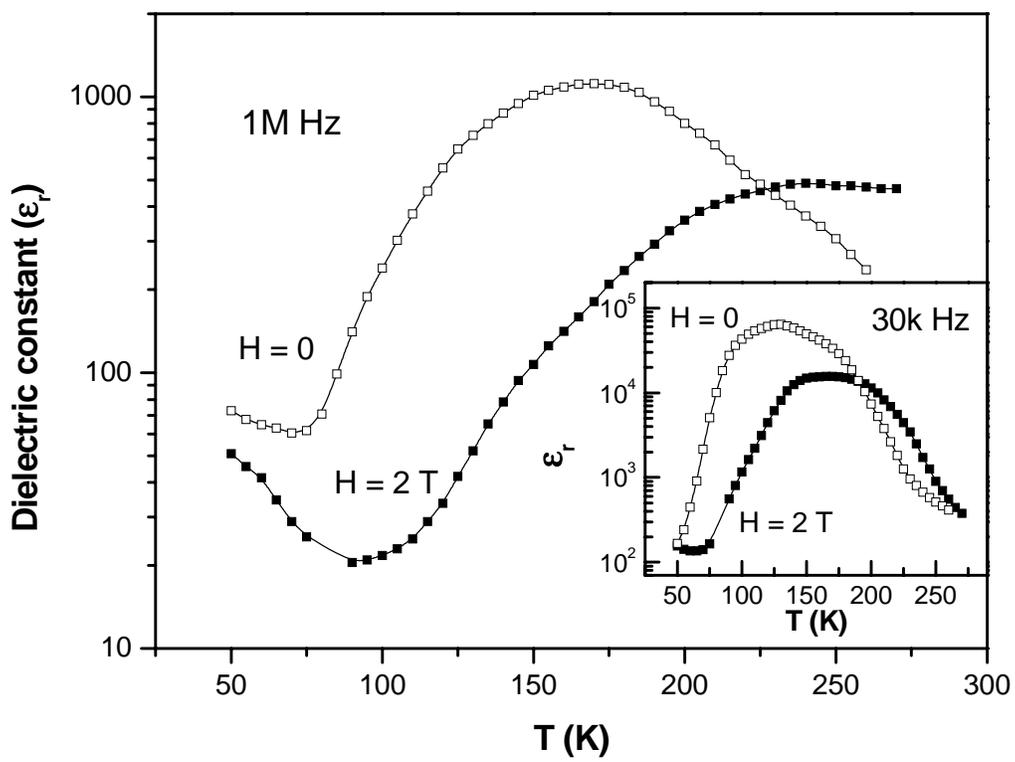

Figure 3

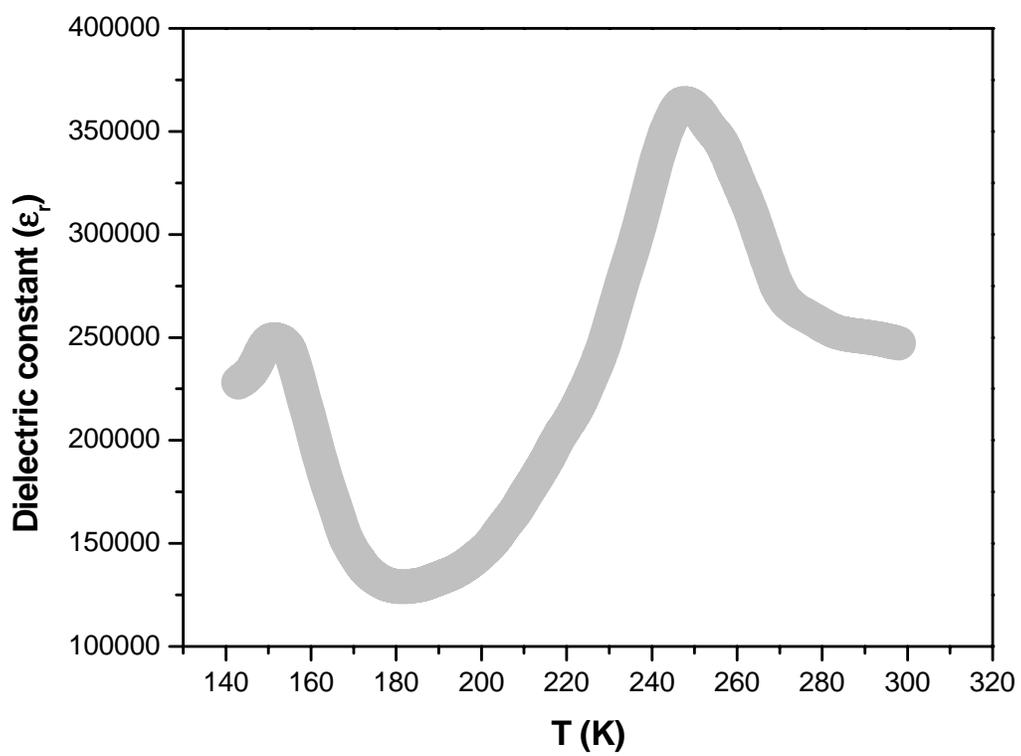

Figure 4

11